# Mössbauer spectroscopic study of a σ-Fe$_{65.9}$V$_{34.1}$ alloy: Curie and Debye temperatures

Jan Żukrowski[1], Stanisław M. Dubiel[2*],

[1]AGH University of Science and Technology, Academic Center for Materials and Nanotechnology, al. A. Mickiewicza 30, 30-059 Kraków, Poland, [2]AGH University of Science and Technology, Faculty of Physics and Applied Computer Science, al. A. Mickiewicza 30, 30-059 Kraków, Poland,

**Abstract**

Sigma-phase Fe$_{65.9}$V$_{34.1}$ alloy was investigated with the Mössbauer spectroscopy. Mössbauer spectra were recorded in the temperature interval of 80-300 K. Their analysis in terms of the hyperfine distribution protocol yielded the average hyperfine field, <B>, the average center shift, <CS>, and the spectral area, A. The magnetic ordering temperature, T$_C$=312.5(5) K was determined from the temperature dependence of <B>, and the Debye temperature, T$_D$, from the temperature dependence of <CS> and the relative spectral area. The value obtained from the former was 403(17) K and that from the latter 374(2) K. The value of the force constant was determined. The lattice dynamics of Fe atoms was described in terms of the kinetic, E$_k$, and the potential energy, E$_p$.

[*] Corresponding author: Stanislaw.Dubiel@fis.agh.edu.pl (S. M. Dubiel)

# 1. Introduction

Sigma ($\sigma$) phase belongs to the Frank-Kasper (FK) family of phases which are also designated as topologically close-packed (TCP) structures. Their characteristic features are high values (12-16) of coordination numbers [1]. An interest in $\sigma$ (and other FK-phases) is twofold: on one hand it has practical reasons, and, on the other hand, scientific ones. The former interest stems from the detrimental effect of $\sigma$ on many useful properties of technologically important materials e. g. steels, super alloys, high entropy alloys. Its presence, even in low percentage, significantly deteriorates useful properties like mechanical strength and resistance to high temperature corrosion e. g. [2]. In these circumstances, the interest in $\sigma$ from the practical viewpoint is rather concentrated on efforts aimed at development of such materials in which its precipitation is suppressed or, at least, retarded. The scientific interest in $\sigma$ is challenged by its complex crystallographic structure (tetragonal unit cell - space group $D^{14}_{4h}$ - $P4_2/mnm$ - with 30 atoms residing on 5 non-equivalent lattice sites) , and interesting physical properties. They can be readily tailored by changing alloy elements and chemical composition thanks to the fact that $\sigma$ can exists in a certain composition range. Concerning magnetic properties of $\sigma$ in the Fe-V system, one of the subjects of the present study, they were revealed in early 1960s [3], and until recently regarded as ferromagnetic. Not long ago, its magnetism was shown to be more complex than believed viz. it had turned out to have a re-entrant character [4]. The Fe-V system is especially interesting with regard of $\sigma$, as the phase can be formed within a wide range of composition viz. ~33 - ~65 at% V [5]. This gives a unique opportunity for changing physical properties of $\sigma$ in a wide range of composition. Concerning the magnetic ones, the magnetic ordering temperature (Curie point), $T_C$, can be continuously raised up to above room temperature. The actual record obtained for the $Fe_{65.6}V_{34.4}$ alloy is as high as ~307K as determined form magnetization measurements and ~324K as found from Mössbauer spectroscopic study [6].

Regarding the Debye temperature, $T_D$, of the $\sigma$-$Fe_{100-x}V_x$ alloys (34.4 $\leq$ x $\leq$ 59) so far it was determined only with the Mössbauer spectroscopy from a temperature dependence of the center shift [7]. The $\theta_D$-values span between ~420 and ~600K and do not show any monotonous behavior.

This paper reports results the Curie and Debye temperatures as determined using the Mössbauer spectroscopy on a σ-Fe$_{65.9}$V$_{34.1}$ sample which, to our best knowledge, is the σ-FeV alloy with the lowest concentration of V investigated so far.

**2. Experimental**

**2.1. Sample preparation and characterization**

Master alloy of a nominal composition α-Fe$_{66.5}$V$_{33.5}$ was prepared by melting 1323 mg of Fe (99.95%purity) and 609 mg of V (99.5%purity) in an arc furnace under protective argon atmosphere. A loss of mass caused by the melting was 2 mg which corresponds to an uncertainty in the concentration of ±0.1 at%. The ingot was flipped over and re melted three times before it was solution treated at 1273K for 72h to increase its homogeneity. Finally, it was quenched onto a block of brass kept at 295K. The transformation into the σ-phase was performed by annealing the solution-treated ingot at T = 973 K for 14 days. The verification of the α-to-σ phase transformation was done by recording room-temperature X-ray pattern which is shown in Fig. 1. Its analysis with the FullProf software gave evidence that the master alloy was transformed into σ to 97.5(5)%. The untransformed residual phase (2.5%) was identified as α-iron. Thus the stoichiometry of σ can be regarded as Fe$_{65.9}$V$_{34.1}$. The parameters of the unit cell of σ were found to be as follows: *a*=8.8674(5) Å and *c* =4.6015(4) Å. They are in line with those found previously for the alloys with higher contents of vanadium [7].

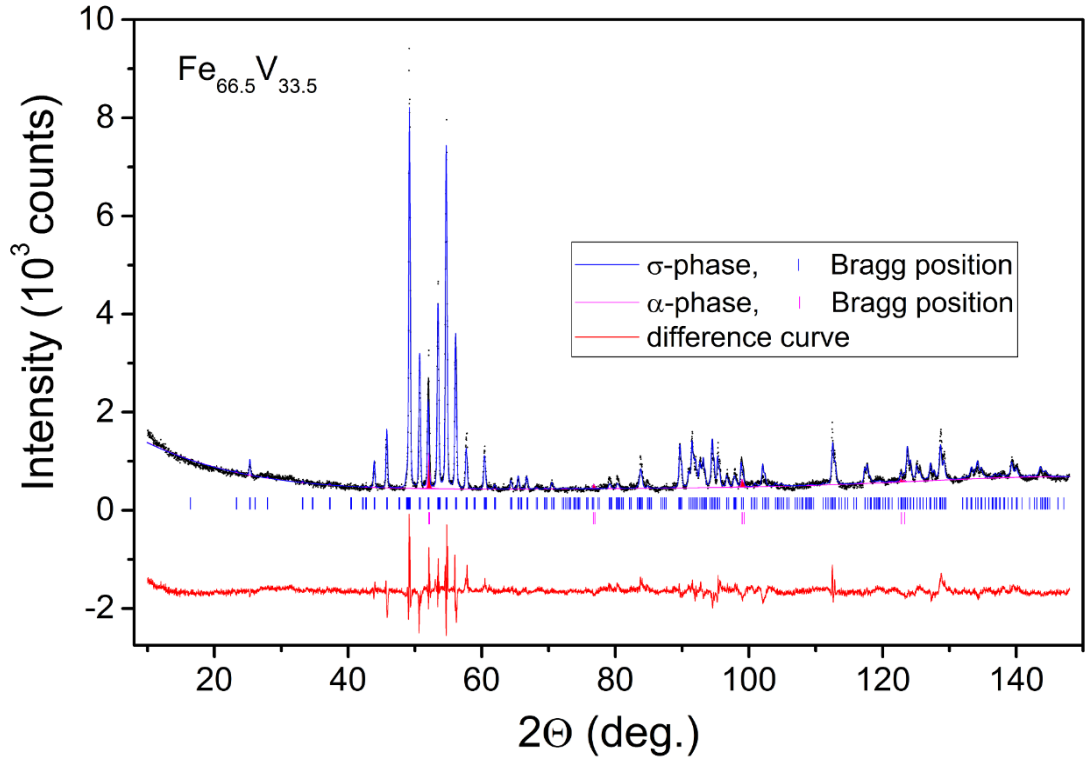

Fig. 1 The Rietveld refinement XRD pattern of Fe$_{66.5}$V$_{33.5}$ alloy recorded at room temperature. Positions of peaks of the σ and α phases are indicated by vertical bars in black and red, respectively. A difference curve between the measured and the calculated patterns is also displayed.

## 2.2. Mössbauer spectra measurements and analysis

Mössbauer spectra were recorded in a temperature range of 80-330K using a standard spectrometer working in a constant acceleration mode. Gamma rays of 14.4 keV were supplied by a $^{57}$Co source embedded into a Rh matrix. Examples of the spectra are presented in Fig. 2 (left panel). They are very similar in shape to those recorded previously on the σ-Fe$_{65.6}$V$_{34.4}$ alloy [6]. The spectra were analyzed to yield spectral parameters pertinent to two quantities of interest viz. the magnetic ordering temperature (Curie), $T_C$, and the Debye temperature, $T_D$. The former can be readily determined from a temperature dependence of the average hyperfine field, *<B>*, while the latter in two ways: from a temperature dependence of (a) the center shift, *CS*, and (b) relative recoil-free fraction, *f/f$_o$*.

Mössbauer spectra were analyzed using a least-square procedure and the transmission integral approach. All three hyperfine interactions were taken into account. Each spectrum was assumed to be composed of five sub spectra following the fact that Fe atoms occupy all five lattice sites in the unit cell of σ. The shape of each sub spectrum was assumed to have the Voight's profile. Relative contributions of the sub spectra, $W_k$, were equal to the corresponding relative lattice site occupancies by Fe atoms as revealed from neutron diffraction experiment [8]. They were kept constant in the fitting procedure. A center shift, $CS_k$, (k=1,2,3,4,5)) of each sub spectrum was a sum of the isomer shift characteristic of a given sub lattice as reported elsewhere [9] and the SOD term. The latter was common to all five lines and treated as a free parameter. Its temperature dependence was next used to determine the Debye temperature and the mean-square velocity of vibrations – section 3.2.

To get the average hyperfine field, <B>, each of the five sub spectra were assumed to have a Gaussian distribution of the field, and the Voigt's profile of the lines was assumed [10]. The average hyperfine field, <B>, was obtained by integrating the Gaussians. Examples of the hyperfine distribution curves are shown in Fig. 2.

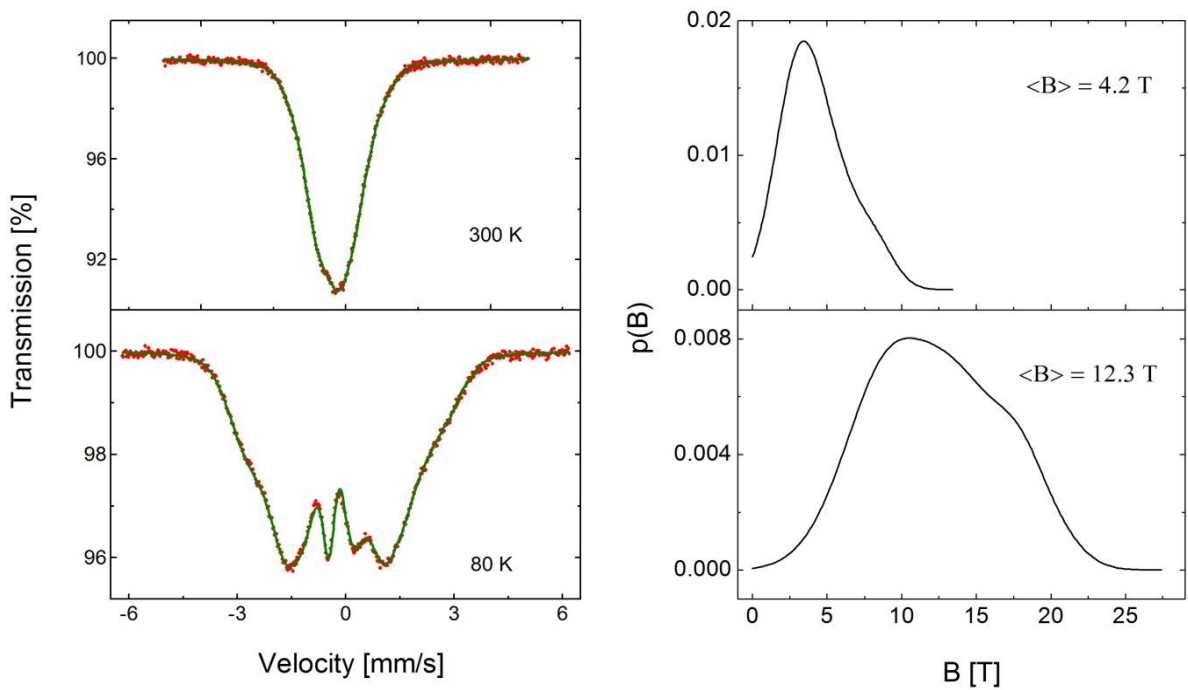

Fig. 2 (left panel) Mössbauer spectra recorded on the σ-Fe$_{65.9}$V$_{34.1}$ alloys at selected temperatures shown, (right panel) corresponding curves of the hyperfine field distribution.

## 3. Results and discussion

### 3.1. Curie temperature

Temperature dependence of <B> is illustrated in Fig. 3. The inset shows all measured data (from which it is evident that the sample is magnetic with the ordering temperature ~313K), while the main body of this figure presents the data in the magnetic phase analyzed in in terms of the Brillouin function. The best fit to the data - marked by the dashed line – yielded the magnetic ordering temperature of $T_C$= 312.4(5) K, and the total electronic angular momentum quantum number, $J$=3. The extrapolated to zero temperature value of <B> = 12.07(5) T is equivalent to the magnetic moment of 0.9 μ$_B$ using the conversion constant reported elsewhere [6]. This means that, contrary to expectation, the magnetism of the presently investigated sample is not evidently stronger than that of the σ-Fe$_{65.6}$V$_{34.4}$ one despite a higher content of Fe in the former. Following the up-to-date known $T_C(x)$ – relationship viz. an increase of $T_C$ ≈10K per at% Fe in the range of ~34.4-40 at% V [6], one would expect $T_C$ ≈325K for the σ-Fe$_{65.9}$V$_{34.1}$ alloy. It should be, however, added that the presently found value of $T_C$ perfectly agrees with the one determined from the AC magnetic susceptibility measurements for the same alloy [11]. The AC susceptibility value of $T_C$ is higher than the corresponding value found from the DC magnetization measurements performed on the σ-Fe$_{65.4}$V$_{34.4}$ [6] i.e. it is in line with the general trend observed for other compositions. It is not clear why the values of $T_C$ determined from the temperature dependence of the average hyperfine field for the two samples with close concentrations viz. x=34.4 and 34.1 differ by ≈10K. Possibly, the difference may have its origin in different site occupations because the α→σ transformation was performed in different conditions i.e. at 973 K for Fe$_{65.4}$V$_{34.4}$ and at 1273 K for Fe$_{65.9}$V$_{34.1}$. This may also be a reason why in the latter case the transformation was not 100% successful.

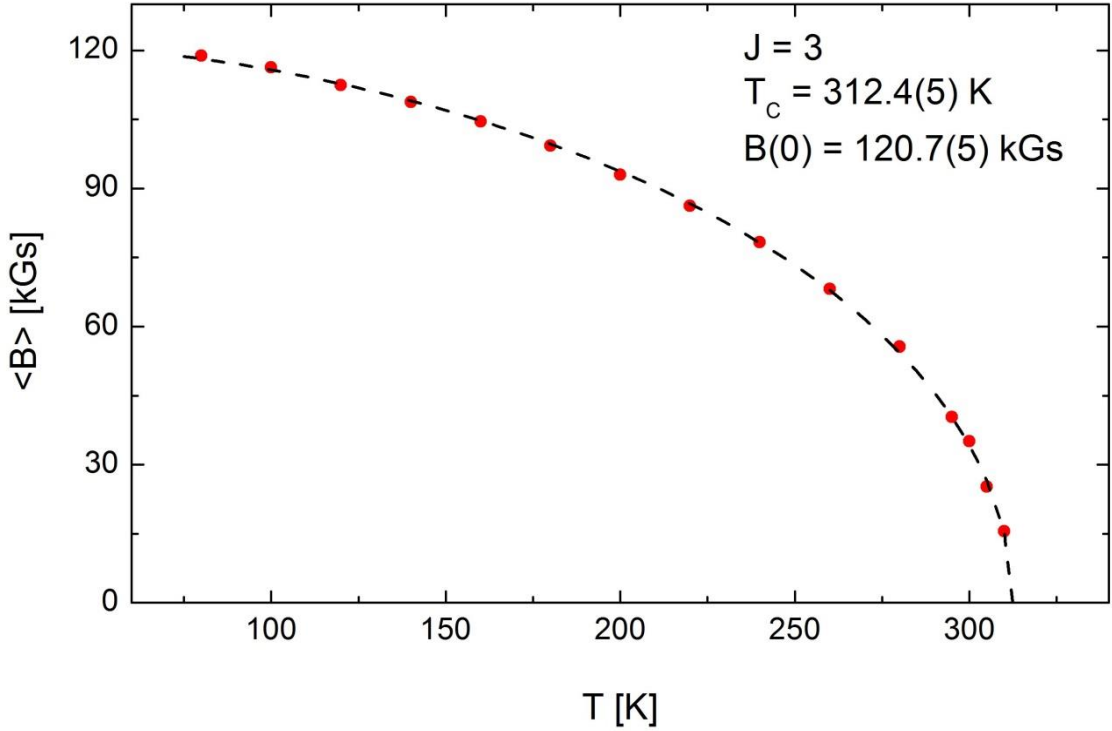

Fig. 3 Average hyperfine field, <B>, vs. temperature, T. The dashed line stands for the best fit of the Brillouin function to the data with the total angular momentum quantum number J=3.

## 3.2. Debye temperature

### 3.2.1. Temperature dependence of center shift

The temperature dependence of the center shift, CS(T), can be expressed by the following equation:

$$CS(T) = IS(0) - \frac{3k_B T}{2mc}\left[\frac{3T_D}{8T} + \left(\frac{T}{T_D}\right)^3 \int_0^{T_D/T} \frac{x^3}{e^x - 1}dx\right] \quad (1)$$

Where *IS(0)* stays for the isomer shift (temperature independent), $k_B$ is the Boltzmann constant, *m* is a mass of $^{57}$Fe atoms.

The second term in eq. (1), known as the second-order Doppler shift, *SOD*, depends on the mean-squared velocity of the vibrating atoms, $<v^2>$, via the following equation:

$$SOD = -\frac{E_\gamma}{2c^2}\langle v^2 \rangle \qquad (2)$$

Where $E_\gamma$ stands for the energy of γ-rays (here 14.4 keV) and *c* is the velocity of light.

The *CS(T)* dependence found in the present study is illustrated in Fig. 4. The best-fit of eq. (1) to the measured data, shown in Fig. 4 as a solid line, yielded $\theta_D$ = 403(17)K.

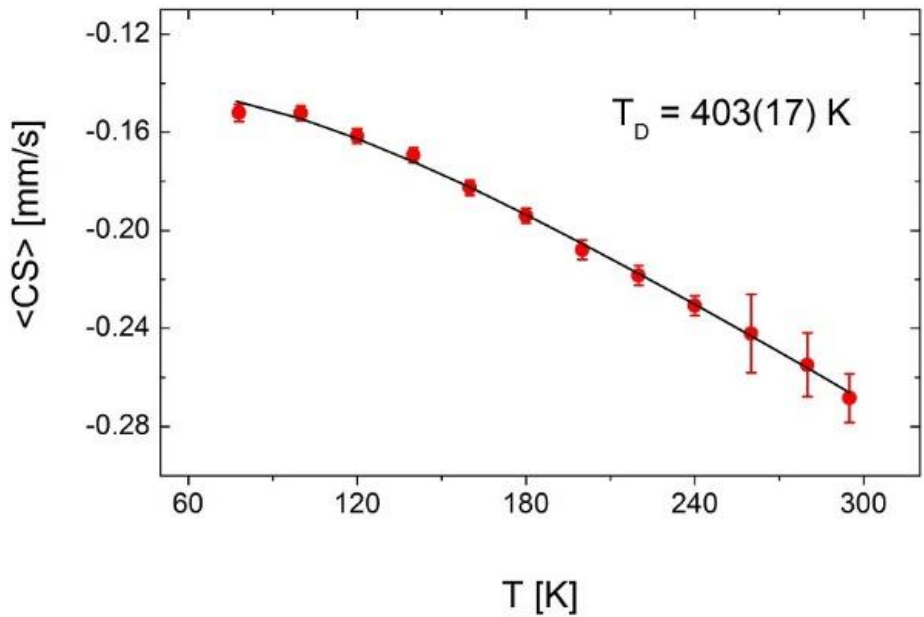

Fig. 4 Temperature dependence of the average center shift, <CS>. The solid line represents the best fit of eq. (1) to the data.

### 3.2.2. Temperature dependence of recoil-free fraction

The recoil-free fraction, f, also known as the f-factor, can be written as follows:

$$f = \exp[-(\frac{E_\gamma}{\hbar c})^2 \langle x^2 \rangle] \quad (3)$$

Where $<x^2>$ is the mean-square amplitude of vibrating atoms. It is related to $\theta_D$ via the following expression:

$$f = \exp[-\frac{6E_R}{k_B T_D}\{\frac{1}{4} + (\frac{T}{T_D})^2 \int_0^{T_D/T} \frac{x}{e^x - 1} dx\}] \quad (4)$$

Where $E_R = \frac{E_\gamma^2}{2mc^2}$ is the recoil energy.

In the approximation of a thin absorber, the f-factor is proportional to a spectral area, A. In practice one uses a normalized spectral area, $A/A_o$, as a measure of the relative f-factor, $f/f_o$ ($A_o$ being the spectral area at the lowest temperature – 80 K in this case). The temperature dependence of $ln(f/f_o)$ is presented in Fig. 5. A good agreement between the experimental data and the theoretical fit is evident, and the resulting value of the Debye temperature is equal to 374(2)K. This value of $T_D$ is significantly lower than the one obtained from the CS(T) relationship. It is, however, known that the value of the Debye temperature depends, in general, on a method used for its determination, and it can vary significantly. Although in the present case we used the same method for determining the Debye temperature it was determined using two different spectral quantities viz. CS and f. They give different information on the lattice dynamics i.e. the former on the mean-squared velocity, hence a kinetic energy, $E_k$, and the latter on the mean-squared displacement, hence on the potential energy, $E_p$, of the vibrating atoms. Consequently, values of $T_D$ as determined from CS and f are usually different. For example, in metallic iron $T_D$ = 421(30) K was found from CS whereas $T_D$ = 358(18) K from f [12].

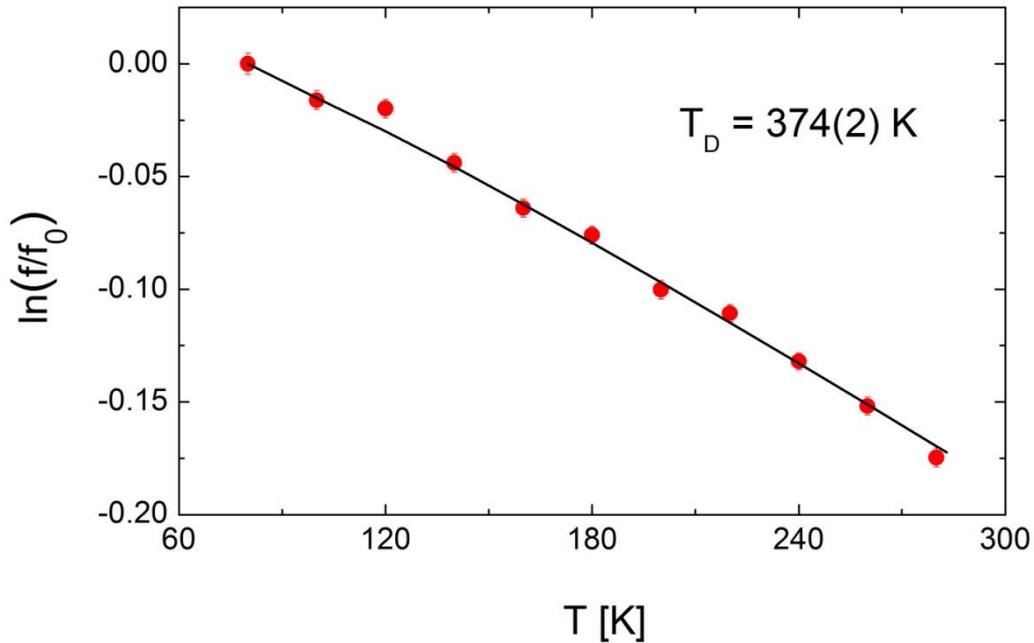

Fig. 5 Temperature dependence of ln(f/f$_o$) and its best fit to eq. (4) shown as a solid line.

### 3.3. Energy relations

#### 3.3.1. $<v^2>$ - $<x^2>$ correlation

As evidenced in Fig. 6 the mean-square velocity of vibrating Fe atoms, $<v^2>$, is linearly correlated with the corresponding mean-square amplitude, $<x^2>$. The force constant, $D$, can be calculated based on this correlation. Namely, $D=m\cdot\alpha$, where $\alpha$ is the slope of the best-fit line in the $<v^2>$ - $<x^2>$ relationship, and $m$ is the mass of $^{57}$Fe atoms. In this way the value of $D$=188(6) N/m was found. It compares reasonably well with the value of 157(2) N/m derived from the partial density of states measured at 298K for a σ-Fe$_{52.5}$Cr$_{47.5}$ alloy [13]. The obtained $D$-value can be next used to calculate the potential energy, $E_p=0.5\cdot D\cdot<x^2>$.

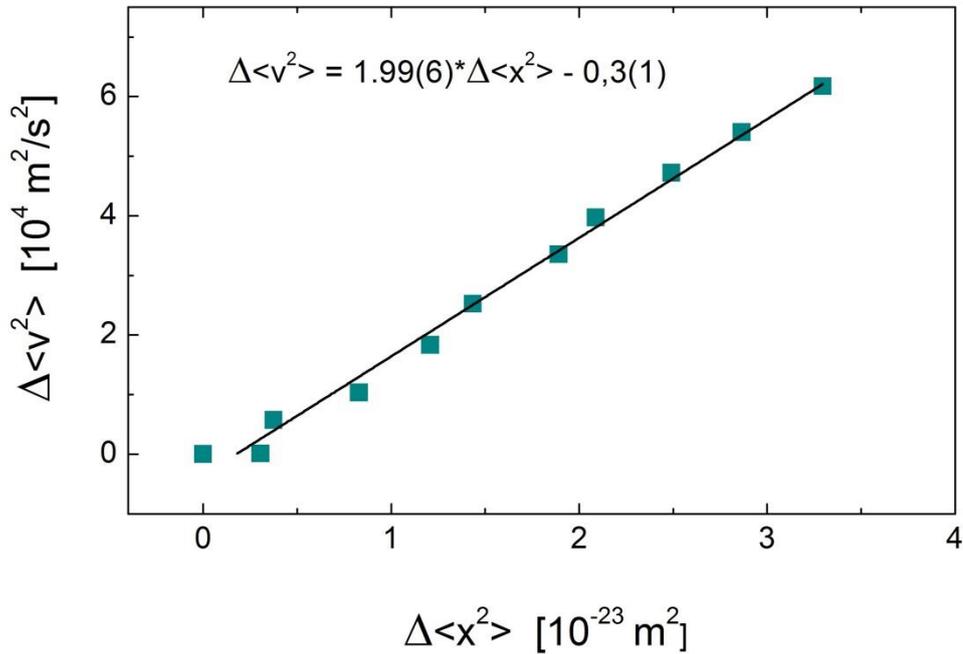

Fig. 6 Relationship between the mean-square velocity of the vibrating Fe atoms, $<v^2>$, and the mean-square amplitude of these vibrations, $<x^2>$. The solid line represents the best linear fit to the data.

### 3.3.2. $<x^2>$ and potential energy

The knowledge of $f/f_o$ enables, via eq. (3), determination of a relative change of the mean-square amplitude of vibrating atoms, $\Delta<x^2> = <x^2> - <x_o^2>$, and, in turn, a relative temperature-induced change of the potential energy, $\Delta E_p = 0.5 \cdot D \cdot \Delta<x^2>$. It dependence on temperature is linear as shown in Fig. 7.

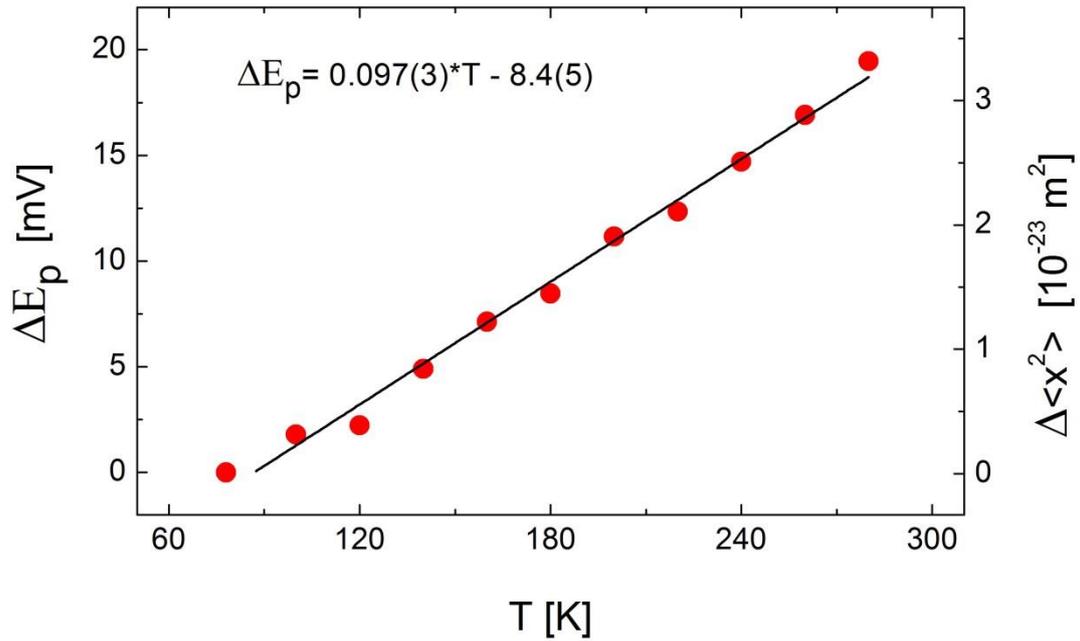

Fig. 7 Relative change of the potential energy, $\Delta E_p$, and of the mean-square amplitude of vibrating atoms, $\Delta <x^2>$, as a function of temperature, $T$. The solid line stands for the best linear fit to the data.

### 3.3.3. $<v^2>$ and kinetic energy

The equation (2) enables determination of the mean-square velocity of the vibrating Fe atoms, $<v^2>$, hence that of the kinetic energy, $E_k = 0.5 \cdot m \cdot <v^2>$. As we want to compare it with the potential energy of the vibrations which, in the present experiment, can be determined only relatively – see the previous paragraph – we will determine a change of the kinetic energy, $\Delta E_k = 0.5 \cdot m \cdot (<v^2> - <v_o^2>)$, relative to its value at the lowest measured temperature, $E_{ok} = 0.5 \cdot m \cdot <v_o^2>$. Such obtained $\Delta E_k$-values as well as the $\Delta <v^2> = (<v^2> - <v_o^2>)$ ones are plotted in Fig. 8. The best-linear fit to the data is shown as well.

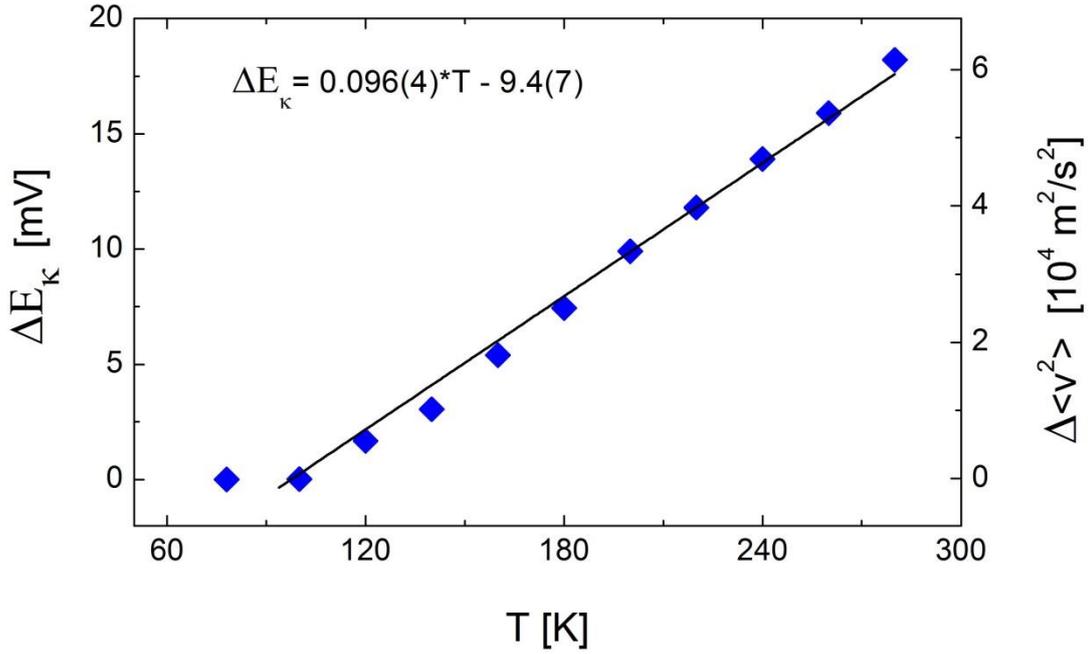

Fig. 8  Relative change of the kinetic energy, $\Delta E_k$, and of the mean-squared velocity of vibrating atoms, $\Delta \langle v^2 \rangle$, as a function of temperature, $T$. The solid line stands for the best linear fit to the data.

### 3.3.4. Kinetic vs. potential energy

The plots of $\Delta E_k$ and $\Delta E_p$ versus temperature show here a very similar and a regular behavior viz. both increase linearly with $T$. This is in contrast to the results found for the $\sigma$-$Fe_{100-x}Cr_x$ (x=46, 48) alloys, where the changes in $E_p$ were up to one order of magnitude higher than the corresponding ones in $E_k$ and their temperature dependence was anomalous [14]. The difference is possibly due to the fact that in the present case the temperature of measurements covers one magnetic phase, while for the $\sigma$-$Fe_{100-x}Cr_x$ alloys the temperature where the anomaly was revealed was much lower (5-35K) and it covered the temperature range in which a reentrant i.e. ferromagnetic-to-spin glass transition occurred. A departure from the harmonic behavior observed in this system could be due to a different degree of a spin-phonon coupling in the ferromagnetic and in the spin glass phases.

## 4. Conclusions

The results obtained in the present study permit the following conclusions to be drawn:

1. The σ-Fe$_{65.9}$V$_{34.1}$ alloy orders magnetically at T$_C$ =312.5(5) K
2. The Debye temperature is equal to 403(17) K as found from the temperature dependence of the center shift, and to 374(23) K as determined from the temperature dependence of the relative spectral area.
3. The force constant responsible for the vibrations of Fe atoms is equal to 188(6) N/m.
4. Both kinetic as well as the potential energy increase linearly with temperature, and the rate of increase is the same which means that the vibrations of Fe atoms are harmonic.

## References

[1] A. K. Sinha, *Topologically Close-packed Structures of Transition Metal Alloys,* Pergamon Press Ltd., Oxford, 1972
[2] M. Venkatraman, K. Pavitra, V. Jana, T. Kachwala, Adv. Mater. Res., 794 (2013) 163
[3] D. Parsons, Nature, 185 (1960) 839
[4] R. Barco, P. Pureur, G. L. F. Fraga, S. M. Dubiel, J. Phys.: Condens. Matter, 24 (2012) 046002
[5] http://www.calphad.com/iron-vanadium.html
[6] J. Cieślak, B. F. O. Costa, S. M. Dubiel et al., J. Magn. Magn. Mater., 321 (2009) 2160
[7] J. Cieslak, M. Reissner, S. M. Dubiel et al., Intermetallics, 18 (2010) 1695
[8] J. Cieślak, M. Reissner, S. M. Dubiel et al., J. Alloys Comp., 460 (2008) 20
[9] J. Cieślak, J. Toboła, S. M. Dubiel, Phys. Rev. B, 81 (2010) 174203
[10] D.G. Rancourt, J.Y. Ping, Nucl. Instrum. Method Phys. Res. B58 (1991) 85
[11] M. Bałanda, S. M. Dubiel, *AC magnetic susceptibility study of a sigma-phase Fe$_{65.9}$V$_{34.1}$ alloy*, http://arxiv.org/abs/1509.03719
[12] G. Chandra, C. Bansal, J. Rey, Phys. Status Solidi (a), 35 (1976) 73
[13] S. M. Dubiel, J. Cieślak, W. Sturhahn et al., Phys. Rev. Lett., 104 (2010) 155503
[14] S. M. Dubiel, J. Cieslak, M. Reissner, EPL, 101 (2013) 16008